\documentclass[reprint,amsmath, aps, amssymb,superscriptaddress,prl]{revtex4-1}
\usepackage{braket}
\usepackage{graphicx}
\usepackage{natbib}
\usepackage{dsfont}
\usepackage{mathtools}
\usepackage{hyperref}

\begin{document}
\title{Quantum simulations of dissipative dynamics: time-dependence instead of size}
\date{\today}
\begin{abstract}
The simulation of quantum systems has been a key aim of quantum technologies for decades, and the generalisation to open systems is necessary to include physically realistic systems. We introduce an approach for quantum simulations of open system \emph{dynamics} in terms of an environment of minimal size and a time-dependent Hamiltonian. This enables the implementation of a continuous-time simulation with a finite environment, whereas state of the art methods require an infinite environment or only match the simulation at discrete times. We find necessary and sufficient conditions for this Hamiltonian to be well behaved and, when these are not met, we show that there exists an approximate Hamiltonian that is, and look into its applications.
\end{abstract}
\author{Benjamin Dive}
\author{Florian Mintert} 
\affiliation{Department of Physics, Imperial College, SW7 2AZ London, UK}
\author{Daniel Burgarth}
\affiliation{Institute of Mathematics, Physics and Computer Science, Aberystwyth University, SY23 3FL Aberystwyth, UK}
\maketitle

\subsection{Introduction}
\label{sec:Introduction}
Every quantum system inevitably interacts with its environment. As quantum simulations are a key aim of quantum technologies \cite{Cirac2012}, the question of how open systems can be simulated efficiently on a quantum computer is one which has received significant research interest in recent years \cite{Koniorczyk2002, Ziman2004, Koniorczyk2005, Kliesch2011, Rybar2012, Sweke2014, Barreiro2011}. In addition to this, open systems have been shown to be useful for state engineering \cite{Carollo2006,Huelga2012}, and as an alternative model of quantum computation \cite{Verstraete2009}. The dynamics of an open quantum system is described by a time-dependent quantum channel, which is a completely-positive trace-preserving map which acts on quantum states in such a way that probabilities stay well defined \cite{Jamiokowski1972, Choi1975, Breuer2002, Caruso2012}. Such channels are often derived by assuming that the system interacts via a Hamiltonian with an environment, which is then traced out because it is not experimentally accessible. A possible way of simulating open systems is therefore to recreate this system-bath interaction in a controllable manner, called a dilation, and implement the dynamics directly \cite{Evans1977}. However, this is rarely feasible due to the infinite size of the environment and the intrinsic difficulty in engineering such a system. Another method is to use Stinespring's theorem \cite{Stinespring1955, Heinosaari2012}, which states that every quantum channel at a given point in time is equivalent to a unitary acting on a larger state followed by tracing out the ancilla, to create a finite dilation. This, however, has the disadvantage of modelling the evolution to a fixed point in time only, rather than replicating the dynamics for all times. Performing a series of Stinespring dilations one after the other allows the evolution to be matched at discrete times, at the cost of a large increase in the ancilla space \cite{Ziman2004, Rybar2012}.

The central idea of this paper is to find, for a time-dependent quantum channel which describes the evolution of a system, a \emph{finite} dilation such that the evolution matches \emph{at all times}, and is smooth so that it gives rise to a well behaved Hamiltonian. Having such a dilation allows the dynamics of an arbitrary open system to be simulated continuously in time simply by acting on a finite system with a physically sensible Hamiltonian. This is useful for quantum simulations, particularly when the time at which the system will be measured is not known beforehand, such as in schemes which rely on photon counting. It is also applicable in cases where we desire to monitor the system continuously via weak measurements \cite{Jacobs2014}. This allows information about the behaviour of the system over an interval of time to be recovered; a situation where prior approaches which evolve the system to a fixed point in time would fail. Furthermore, by shifting the complexity from infinite space to time-dependence, it provides a model on which open systems can be studied easily \cite{Arenz2014}, which we use to investigate how the system-environment interaction is affected by adding controls to the system.

It is known that given two channels which are close by it is possible to find two unitary dilations which are also close by \cite{Kretschmann2008}; we add the stronger constraint of the unitaries varying smoothly so that we can define a Hamiltonian for the dilation. We provide an explicit method for constructing such dilations, and establish precise relations between the continuity and boundedness of this Hamiltonian and properties of the original quantum channel. We show that it is always possible to find such a Hamiltonian which matches the dynamics arbitrarily well, provided that the original evolution is analytic in time. Although our methods are not limited to Markovian channels, we pay particular attention to these as they constitute some of the most common types of noise encountered in quantum information \cite{Rivas2014}. These are the class of channels which are memoryless so that they can be expressed in terms of an equation of motion, $\dot{\rho} = L_t(\rho)$, where $L_t$ is a Lindbladian \cite{Kossakowski1972, Lindblad1976, Gorini1976, Breuer2002}. This allows us to understand this work as raising dilations to the levels of generators; mapping Lindbladians into Hamiltonians. More generally, we study the usefulness of these dilations in the simulation of open systems, give several explicit examples for common quantum channels, and look at the effect of adding Hamiltonian controls; as well as how to generalise the approach to systems consisting of many qubits.

\subsection{Method}
\begin{figure}[h]
\includegraphics[trim = 0.5cm 4cm 2cm 2.5cm, width=0.95\columnwidth]{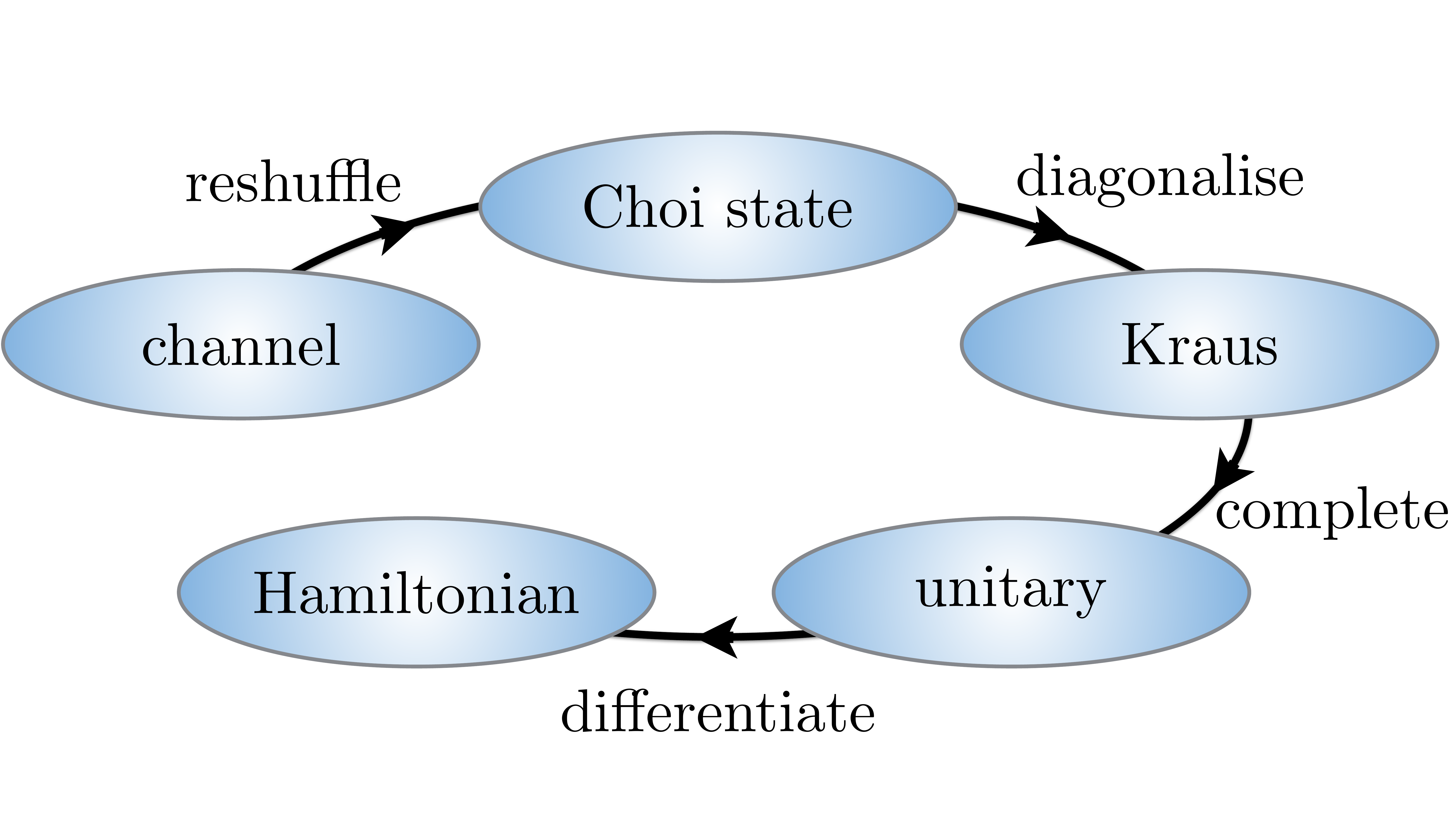}
\caption{A schematic of how to construct the dilation from a quantum channel. All the steps preserve the analyticity in time apart from separating the diagonalised Choi state into Kraus operators which introduces a square root. It is differentiating this square root that introduces the possibility of discontinuities and divergences.}
\label{flowchart}
\end{figure}

We begin by presenting a series of steps, illustrated in FIG. \ref{flowchart}, which allows us to construct a dilation. Although the Stinespring dilation is highly non-unique, we use this method for two reasons. Firstly, it gives us an explicit way of constructing a unitary dilation \cite{Havel2002, Andersson2007, Heinosaari2012}. Secondly, and for our purposes crucially, it enables us to study the properties of the Hamiltonian by letting us follow the time-dependency of the objects throughout their transformations.

We start with a family of quantum channels, $\epsilon_t(\cdot)$, which is analytic in time. In the case of Markovian channels, these are generated by a Lindbladian according to $\epsilon_t (\cdot) = \mathcal{T}\; \text{exp}\left({\int_0^t d\tau L_\tau(\cdot)}\right)$ where $\mathcal{T}$ is the time-ordering operator, but we do not limit ourselves to such cases. The next step is to construct the (unnormalised) Choi state which arises from the Choi-Jamio\l kowski isomorphism \cite{Heinosaari2012}, and is given by $\Lambda(t) = (\epsilon_t \otimes \mathcal{I})\ket{\Omega}\bra{\Omega}$ where $\ket{\Omega} = \sum_j^d \ket{j j}$ is a maximally entangled state between the system's Hilbert space and its duplicate, and $\mathcal{I}$ is the identity map. This is equivalent to reshuffling the elements of the channel represented as a matrix \cite{Bengtsson2006}. This is just a linear transformation, therefore $\Lambda(t)$ is also analytic in $t$. As $\Lambda(t)$ is Hermitian and positive, it can be decomposed into its eigenvalues and eigenvectors, $\lambda_k(t)$ and $\ket{v_k(t)}$ respectively, where the index $k$ runs from $1$ to the Kraus rank of the channel, $R$, which is upper bounded by $d^2$. It is known via perturbation theory that, as $\Lambda(t)$ is Hermitian and analytic, its eigenvalues and vectors are also analytic for real $t$ \cite{Kato1980}. This allows us to write the channel in its Kraus representation, $\epsilon_t(\rho) = \sum M_k(t) \,\rho\, M_k^\dagger(t)$, where the Kraus operators are \begin{equation}
\label{eq:KrausDefine}
M_k(t) = \sum_{ij} \sqrt{\lambda_k(t)} \braket{i\otimes\Omega| v_k(t)\otimes j} \ket{i}\bra{j}
\end{equation}
for any choice of basis $\{\ket{i}\}$. The presence of the square root here is key. It results in the Kraus operators being continuous everywhere, and smooth everywhere apart possibly from individual points.

The final step in finding a dilation is solving $\sum M_k(t) \,\rho\, M_k^\dagger(t) = \text{Tr}_B [U(t) \,\rho_A\otimes\omega_B\, U^\dagger(t)]$, where we denote the system space with the subscript $A$ and the ancilla with $B$. We are free to choose the initial ancilla state and so pick it to be $\omega = \ket{0}\bra{0}$, which gives as a solution
\begin{equation}
\label{eq:KrausConstraint}
\bra{k_B}U(t)\ket{0_B} = M_k(t) \;\;\forall\;k.
\end{equation}
This provides a dilation where the dimension of the ancilla is $R$, and is known to be the smallest sized ancilla which may be required. The dilation unitary has dimensions $Rd\times Rd$ and Eq.(\ref{eq:KrausConstraint}) constrains $d$ of its columns. These can be thought of as forming $d$ orthonormal vectors in an $Rd$ dimensional space. As we also need to ensure that the dilation is unitary, we impose that $U(t)U^\dagger(t) = \mathds{I}$ which is equivalent to requiring that the remaining $d(R-1)$ columns complete the orthonormal space. We prove in appendix A that this can always be done in a smooth way whenever the $M_k(t)$ are smooth. The desired Hamiltonian satisfies the Schr\"odinger equation, and is therefore given by $H(t) = i\,\dot{U}(t) U^\dagger(t)$.

The Hamiltonian involves the derivative of $U$, which is continuous but in general not analytic in $t$, thus there is the possibility of it being discontinuous or divergent due to the behaviour of the derivative of $\sqrt{\lambda(t)}$. A careful analysis shows two potential problems. Firstly, $\dot{M}_k(t)$ can be discontinuous (but always bounded) when $\lambda_k(t\ne0) = 0$. Secondly, it diverges when $\lambda_k(0) = 0$ but $\dot{\lambda}_k(0) \ne 0$. These properties are inherited by the Hamiltonian, bar some accidental cancellation. The first case, where the Hamiltonian has a step change at some later time, corresponds to the Kraus rank of the channel decreasing at a single point in time. More interesting is the second case. The divergence of the dilation Hamiltonian at $t=0$, is avoided if and only if the dissipative part of $\dot{\epsilon}(t=0)$ vanishes (see appendix B for proof). This gives the immediate corollary that all non-trivial time-independent Markovian channels lead to a divergent dilation. Such channels necessarily have that the survival probability of certain states decays linearly at short times. However, we know from the quantum Zeno effect that the survival probability of a state in any unitary system with a bounded Hamiltonian must decay quadratically for short times \cite{Pascazio2014}. This linear decay might be a signature of the unbounded system environment interaction, but often it is merely a consequence of approximations, like infinitely fast relaxation within the environment \cite{Breuer2002}, made in the derivation of the Master equation. In both cases, the divergence of the dilation is inherited from this and is an indicator of Markovianity.

Both the discontinuities and the divergence are, when they happen, benign. The divergence at $t=0$ has only a finite impact on the dynamics as the dilation can always be picked such that $U\to\mathds{I}$ as $t\to0$. This implies that we can approximate the evolution arbitrarily well by replacing the exact Hamiltonian by a bounded one. This can be seen from the relation between the error in the unitary (which is the error in the evolution) and that of the Hamiltonian \cite{Nielsen2006}:
\begin{equation}
||U(t)-U_T(t)|| \le \int_0^t ||H(t') - H_T(t')|| dt',
\end{equation}
where $||\cdot||$ is the operator norm and $H_T$ and $U_T$ are the target Hamiltonian and unitary respectively. We note that the right hand side is bounded by $\int_0^t ||H_T(t')|| dt'$ for any reasonable applied Hamiltonian, and is always finite (even if $H_T(0)$ is not). Thus the error in the unitary is always finite and can be decreased arbitrarily by having the applied Hamiltonian differ from the exact Hamiltonian for a sufficiently short time. We calculate this error for a specific dilation later, and show that it can indeed be made arbitrarily small with ease. In the case of bounded discontinuities, the dilation Hamiltonian itself can be arbitrary well approximated by a continuous one, which leads to the evolution being arbitrarily well approximated. This gives the following result:
\begin{quote}\emph{For an analytic family of quantum channels acting on $d$ dimensional states there always exists a continuous and bounded Hamiltonian acting on at most $d^3$ dimensional states such that the dynamics on the reduced system are arbitrarily well matched at all times.}
\end{quote}

\subsection{Further Methods}

The method we have just discussed works well, but it requires us to diagonalise the Choi state and, in the case that we are starting from an equation of motion, to calculate the channel. In practice, one or both of these may be very difficult to do analytically. Indeed, systems where these are hard to do are the ones we most want to find a dilation for and be able to simulate on a quantum computer, as they are precisely those which are difficult to simulate classically. We therefore present three alternate methods to construct dilations for complicated systems which rely on having found a dilation for a simpler system. The methods are: changing frames, separating into commuting parts, and perturbation theory.

Firstly, we consider how the change of frame of the initial problem translates into a change in the dilation Hamiltonian. We start with the equation of motion in the given frame:
\begin{equation}
\frac{d \rho}{dt} = L_t (\rho).
\end{equation}
This can be represented in a different frame, for example in the interaction picture, by the transformation $\tilde{\rho}(t) = U_0(t)^\dagger\rho(t) U_0(t)$ where $U_0(t) = e^{-i H_0 t}$. The equation of motion in this frame is given by:
\begin{align}
\frac{d \tilde{\rho}}{dt} &=  \tilde{L}_t (\tilde{\rho})\\
\text{where:}\;\;\;\tilde{L}_t (\cdot) &= U_0^\dagger L_t (U_0\cdot U_0^\dagger)U_0 + i [H_0,\,\cdot] .\nonumber
\end{align}
We can relate the dilation of this Lindbladian, with unitary $\tilde{U}(t)_{AB}$ and Hamiltonian $\tilde{H}_{AB}(t)$, to the dilation of the original Lindbladian. We first note that the respective channels obey
\begin{equation}
\epsilon_t(\cdot) = U_0(t) \tilde{\epsilon_t}(\cdot) U_0^\dagger(t)
\end{equation}
as $\tilde{\rho}(0) = \rho(0)$.
This implies that
\begin{align}
\epsilon_t(\rho) &= U_0 \text{Tr}_B \left[ \tilde{U}_{AB} \;\rho\otimes\omega\;\tilde{U}^\dagger_{AB} \right] U_0^\dagger \nonumber\\
&\equiv \left[ U_{AB} \;\rho\otimes\omega\; U^\dagger_{AB} \right]
\end{align}
which directly shows us that the dilation in the original frame is given by $U_{AB} = (U_0 \otimes\mathds{I}) \tilde{U}_{AB}$, and so the Hamiltonian reads
\begin{align}
\label{eq:changeframedilation}
H_{AB} &= H_0\otimes\mathds{I} + (U_0\otimes\mathds{I}) \tilde{H}_{AB} (U_0^\dagger \otimes \mathds{I}).
\end{align}
This gives us a very simple relation between a dilation Hamiltonian in one frame, and a dilation in a different frame. Indeed, the relation is the same as when we change frame in normal unitary dynamics with the small additional step of transforming $U_0 \to U_0\otimes\mathds{I}$.

Secondly, we show how a dilations can  be calculated by separating the Lindbladian into commuting parts, at the cost of increasing the ancilla space. Take two different channels which commute at all times such that
\begin{align}
\epsilon^{(12)}_t(\cdot) &= \epsilon^{(1)}_t\left( \epsilon^{(2)}_t(\cdot)\right). \label{eq:commDil}
\end{align}
By performing the dilations one after the other, and on different ancilla spaces, we have that
\begin{align}
\epsilon^{(12)}_t(\rho) &= \epsilon^{(1)}\left(\text{Tr}_B\left[U^{(2)}_{AB}\,\rho_A\otimes\omega^{(2)}_B\,U^{(2)\dagger}_{AB}\right]\right)\\
&= \text{Tr}_{BC}\left[U^{(1)}_{AC}U^{(2)}_{AB}\,\rho_A\otimes\omega^{(2)}_B\otimes\omega^{(1)}_C\,U^{(2)\dagger}_{AB}U^{(1)\dagger}_{AC}\right], \nonumber
\end{align}
where $U_{AC} \equiv U_{AC}\otimes\mathds{I}_B$ and similarly for the other operators. This gives as the Hamiltonian
\begin{equation}
H^{(12)}_{ABC} =H^{(1)}_{AC}+ U^{(1)}_{AC} H^{(2)}_{AB} U^{(1)\dagger}_{AC}.
\end{equation}
Due to the commutativity, we can inverse the order in Eq.(\ref{eq:commDil}) and obtain
\begin{equation}
H^{(21)}_{ABC} =H^{(2)}_{AB}+ U^{(2)}_{AB} H^{(1)}_{AC} U^{(2)\dagger}_{AB}
\end{equation}
which is, in general, a different Hamiltonian but leads to the same dynamics on the reduced system.

Thirdly, we can use a perturbative approach. We take the Lindbladian for the system to be $L^{(0)}_t + \delta L^{(1)}_t$ with $\delta\ll1$, where we assume we have already found a dilation for $L^{(0)}_t$, and find a new dilation which gives the correct dynamics to first order in $\delta$. The quantum channel for such a Lindbladian is
\begin{align}
\epsilon_t &= \epsilon^{(0)}_t + \delta \int_0^t \epsilon^{(0)}_{(t,\tau)} L^{(1)}_t \epsilon^{(0)}_{(\tau,0)} d\tau + O(\delta^2)\nonumber\\
&\approx \epsilon^{(0)}_t + \delta \epsilon^{(1)}_t
\end{align}
where $\epsilon^{(0)}_{(t_2,t_1)} =  \mathcal{T} e^{\int_{t_1}^{t_2}L^{(0)}_\tau d\tau}$. Constructing the Choi state from the channel is a linear transformation which can be done separately for $\epsilon^{(0)}_t$ and $\epsilon^{(1)}_t$. The eigenvalues and vectors of $\Lambda^{(0)}_t + \delta\Lambda^{(1)}_t$ can be found to first order in $\delta$ using standard methods from perturbation theory. This is much easier to do than diagnosing the Choi state exactly, although some of the advantage is lost if the original state had a high degree of degeneracy which is broken. Having done this, we can easily find the Kraus operators by expanding Eq.(1) in the main body to first order in $\delta$. It is worth noting that if $M^{(0)}_k=0$ it is sufficient to find $M_k$ to $O(\sqrt{\delta})$, as the equation of motion is quadratic in the Kraus operators and we are only interested in finding the dynamics up to first order in $\delta$. In the case that the Kraus rank of the channel is unaffected by the perturbation, so that the above condition does not hold, we can write $M_k = M_k^{(0)} + \delta M_k^{(1)} + O(\delta^2)$. In that case the correction to the dilation unitary satisfies
\begin{align}
\label{eq:KrausPertubation}
\bra{k_B}U^{(1)}(t)\ket{0_B} &= M_k^{(1)}(t) \;\;\forall\;k.\nonumber\\
U^{(1)}U^{(0)\dagger} + U^{(0)}U^{(1)\dagger} &= 0.
\end{align}
The last expression is equivalent to requiring the unitarity condition to hold to first order in $\delta$. This reduces the problem of finding the dilation unitary to a system of linear equations. The Hamiltonian is then given by
\begin{equation}
\label{eq:HamilPertubation}
H^{(1)} = \left(i\frac{dU^{(1)}}{dt} - H^{(0)}U^{(1)}\right) U^{(0)\dagger}
\end{equation}
such that $H = H^{(0)} + \delta H^{(1)} + O(\delta^2)$. In the case that the Kraus rank of the channel does change, similar expressions can be found, although care must be taken to ensure that both terms $O(\sqrt{\delta})$ and $O(\delta)$ are properly accounted for. This perturbative method can be extended to take into account second, or higher, order effects.

\subsection{Examples}

\begin{figure}[h!]
\includegraphics[width=0.9\columnwidth]{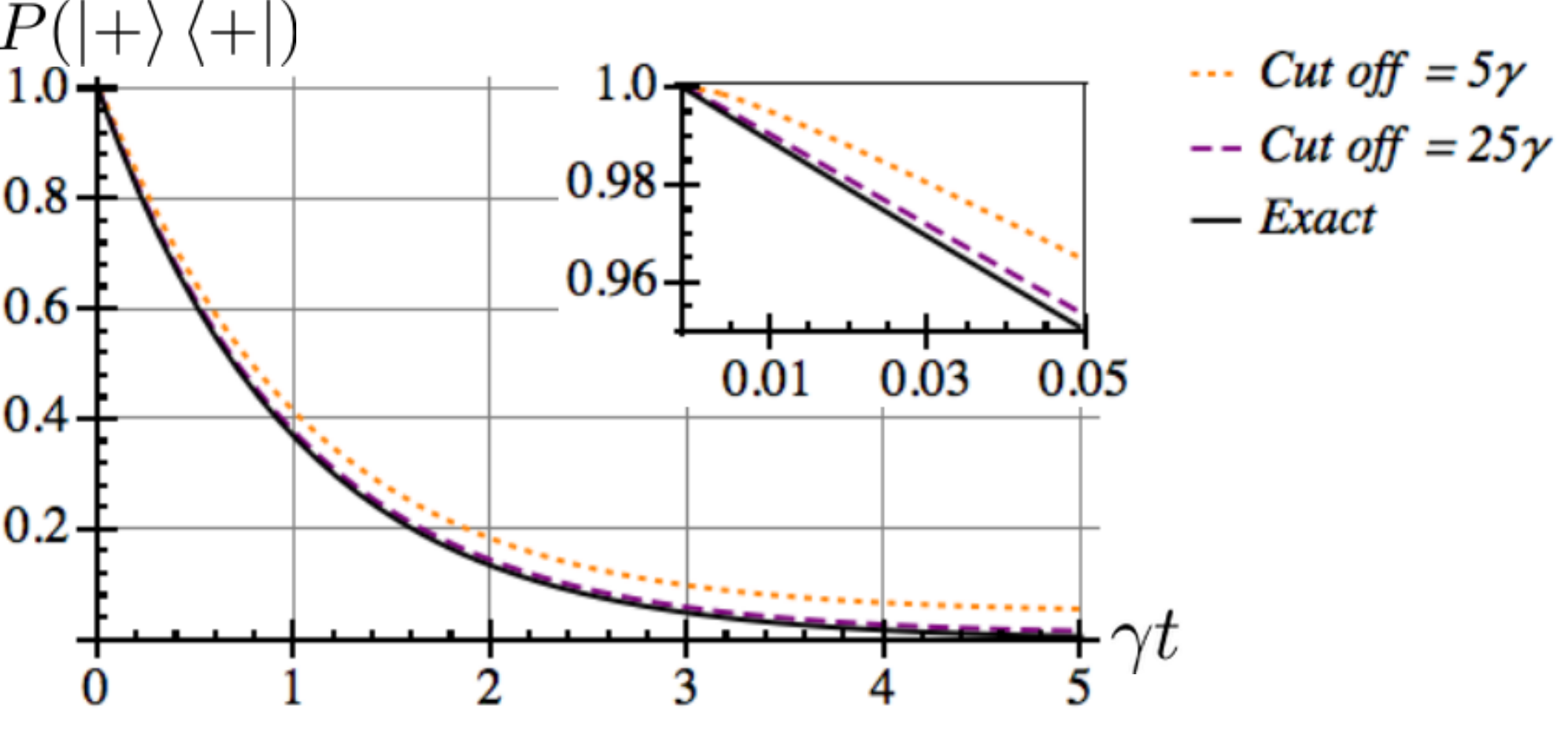}
\caption{The decay of the $\ket{+}\bra{+}$ state when subjected to a dilation of the dephasing channel, where the divergent Hamiltonian is replaced by a finite cut-off at short times. The inset shows the behaviour for short times.}
\label{DecayErrorWITHInset}
\end{figure}

We demonstrate these methods and results by looking at specific examples of quantum channel acting on qubits which represent some of the iconic decoherence models in quantum information \cite{Nielsen2000, Breuer2002, Ladd2010, Franco2013, Rivas2014}. In many cases the resulting Hamiltonians are sufficiently simple to be directly experimentally realisable. Consider the spin-boson model, where a single spin interacts with a bath of bosons via a constant Hamiltonian, such that the reduced dynamics are given by $\dot{\rho}= -\gamma(t) [\sigma_z, [\sigma_z, \rho]\, ]$, where the decay rate is a function of time \cite{Doll2007, Pernice2011}. Performing the steps outlined above, we find the dilation to be
\begin{equation}
\label{eq:SpinBoson}
H_{s.b.}(t) = \frac{\gamma(t)}{2\sqrt{e^{2\int_0^t \gamma(t')dt'} - 1}} \sigma_z \otimes \sigma_y,
\end{equation}
where $\sigma_z$ acts on the system and $\sigma_y$ acts on the ancilla (which consists of a single qubit). The exact form of the Hamiltonian depends on $\gamma(t)$; for a typical spectral density we have that $\gamma(0)=0$ and that, for certain values of $t$, it becomes negative. In such a case the channel is non-Markovian and the dilation is bounded and continuous for all times. In the case of constant $\gamma$, however, this is a Markovian dephasing Lindbladian which gives rise to the phase-flip channel, and whose dilation diverges at $t=0$ as expected.  If we approximate this Hamiltonian to be constant by replacing the scalar pre-factor by $C$ for short times (that is, when $\tfrac{\gamma}{2\sqrt{e^{2\gamma t}-1}}>C$) the error between the target unitary and the unitary reached is upper bounded by $\tfrac{\gamma}{8C}+O(\tfrac{\gamma^2}{C^2})$. Thus, with a sufficiently large $C$ the error can be made arbitrarily small. Another way of looking at the errors is to see how the dynamics of a state depends on $C$, as is plotted in FIG.(\ref{DecayErrorWITHInset}). We see that the main effect of introducing a cut-off is to make the behaviour quadratic, rather than linear, at short time, and that even a modest value for the cut-off is enough to reduce errors to insignificance.

As another example, we look at the dilation for an amplitude damped system \cite{Arenz2014}. Here a non-degenerate, two-level system relaxes to its ground state in a process such as spontaneous emission. We start with an equation of motion $\dot{\rho}= -\gamma(\{\sigma_+\sigma_-, \rho\} - 2\sigma_-\rho\sigma_+) + \tfrac{\omega_0}{2}\sigma_z$, where $\sigma_\pm$ is the raising/lowering operator, and find the dilation to be
\begin{equation}
\label{eq:ampdampdilation}
H_{a.d.}(t) = \frac{i \gamma}{\sqrt{e^{2 t \gamma }-1}} (\sigma_-\otimes\sigma_+ - \sigma_+\otimes\sigma_-) + \tfrac{1}{2}\omega_0\,\sigma_z\otimes\mathds{I},
\end{equation}
where once again the ancilla consists of a single qubit. It is interesting to note that, although this has a very different physical origin to the dephasing channel, the time dependency is almost identical. The comparison highlights some of the most common features of dilations of simple systems. The Hamiltonian is strongest at $t=0$ (possibly even diverging, as discussed previously), and the terms corresponding to decay fall to $0$ for large $t$; which is to be expected for the system to settle to its steady state.

As a more involved example, we now consider a qubit acted on by an amplitude damping Lindbladian as above but with an additional constant driving term, $-i\Omega[\sigma_x,\,\rho]$. In this case the coherent and incoherent part of the evolution no longer commute, which makes a direct calculation of the exact dynamics difficult. Nevertheless, by taking the limit where the driving strength $\Omega$ is much smaller than the decay rate $\gamma$, we can use perturbative methods to find the dilation
\begin{equation}
\label{eq:drivenDila}
H(t) = H_{a.d.}(t) + \Omega\frac{2}{1+e^{\gamma t}} \sigma_x\otimes\mathds{I} +\Omega \frac{\sqrt{e^{2 \gamma  t}-1}}{\left(e^{\gamma  t}+1\right)^2} \sigma_z\otimes\sigma_x,
\end{equation}
where we have set $\omega_0=0$ for simplicity. This dilation has two potentially unexpected features. Firstly, the driving term has gone from being constant to decaying in time. For large $t$ it does this at same rate as $H_{a.d.}$, which is necessary for the system to settle to a single fixed point. Secondly, we have the appearances of a third term, which is back action caused by dilating the control and it has a more complex structure in time, although it too decays at the same rate for large $t$. This term is caused by $H_{a.d.}(t)$ building up entanglement between the system and the ancilla. The emergence of complicated time structure induced by a simple control field is closely linked to the fact that Master equations are changed in a non-trivial way by the addition of a Hamiltonian acting on one subsystem \cite{Dalessandro2014}.

\begin{figure}[t]
\includegraphics[width=0.9\columnwidth]{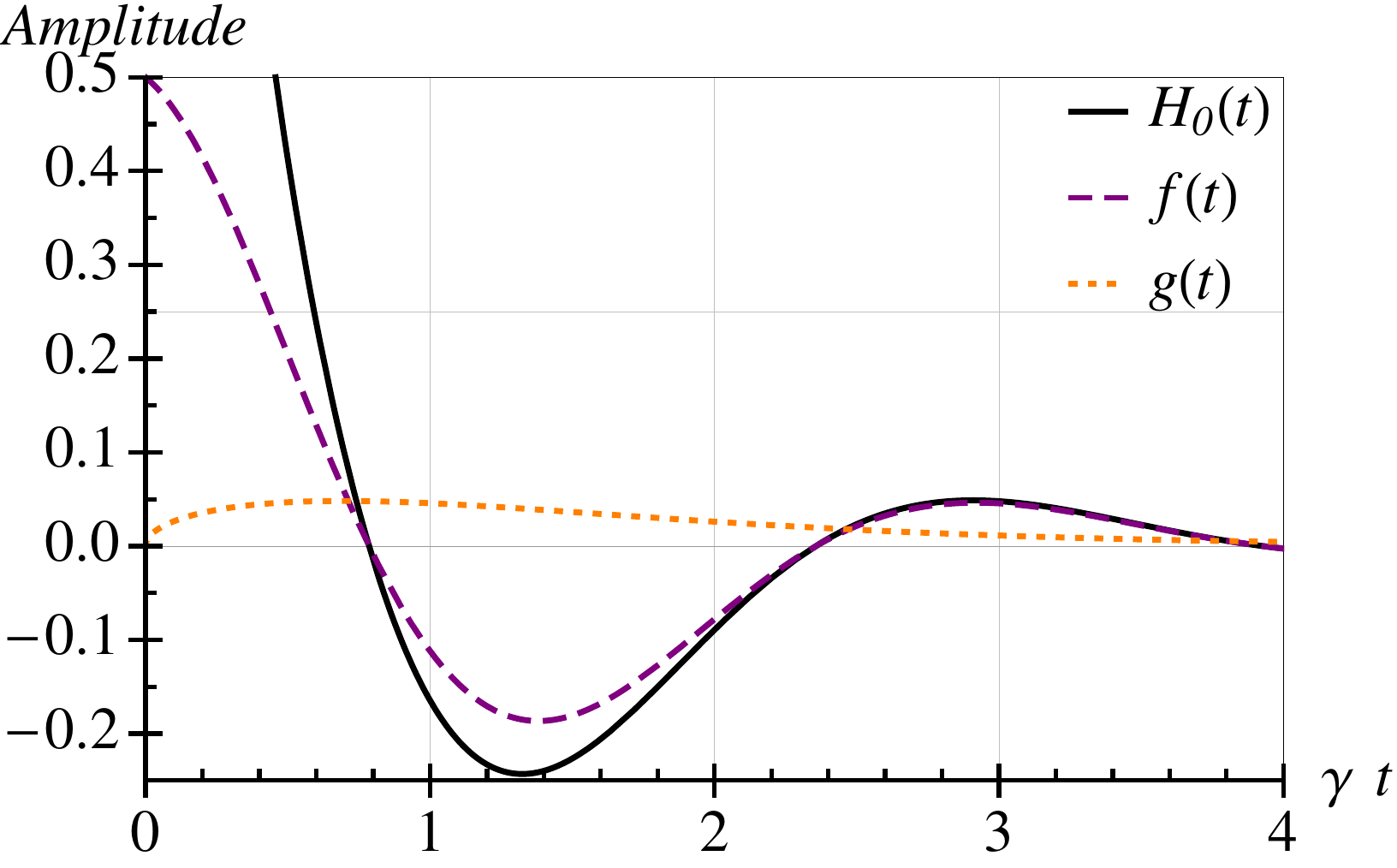}
\caption{The dilation for a qubit subjected to an amplitude damping Lindbladian and a resonant sinusoidal driving field is detailed in Eq.(\ref{eq:RWAHamiltonian}). We plot here the real part of the time-dependent functions in that Hamiltonian with $\omega_0/\gamma=2$. Only one of the terms diverges at $t=0$, while all the terms decay to $0$ quickly at large times.}
\label{fig:RWAHamiltonian}
\end{figure}

A more complex case involving instead a time dependent driving term, $-i\Omega\cos(\omega t)[\sigma_x,\rho]$, can also be dilated in a perturbative method. In this case we also make the Rotating Wave Approximation and, in the resonant case, the dilation is:
\begin{align}
\label{eq:RWAHamiltonian}
H(t) &=  i H_0(t)\,\sigma_-\otimes\sigma_+ + \frac{\omega_0}{4} \sigma_z\otimes\mathds{I} \\
&+ \Omega\,f(t)\,\sigma_-\otimes\mathds{1} +\Omega\,g(t)\,\sigma_z\otimes\sigma_x + h.c. \nonumber
\end{align}
where $H_0(t) = e^{-i\omega_0 t}\gamma/{\sqrt{e^{2 t \gamma }-1}}$, $f(t) = e^{-i \omega_0 t}/({1+e^{\gamma t}})$, and $g(t) = \sqrt{e^{2 \gamma  t}-1}/({4\left(e^{\gamma  t}+1\right)^2})$ are plotted in FIG.\ref{fig:RWAHamiltonian}. The increase in complexity of the Hamiltonian is directly related to the time-dependence of the original equation of motion, but the dilation could still be constructed which shows that this approach is applicable to a wide range of problems.

\subsection{Conclusion}

By rescaling time in a nonlinear way, some of these channels can even be dilated to constant Hamiltonians. In general, this is true whenever the dilation Hamiltonian is of the form $H(t) = h(t) X$ such that it has a single time dependent factor. This allows us to apply a constant Hamiltonian $H = h_0 X$ for time $\tau$ to simulate the real dynamics evolving for a time $t$ where $\tau = \tfrac{1}{h_0}\int_0^t h(t')dt'$. As $h(t)$ is continuous and bounded, this is always well defined. In cases where there are several different time dependencies, this method can be used to remove one of them. This is particularly useful in eliminating divergences, which would otherwise be problematic to implement experimentally. It also has the advantage that, in many cases, the evolution for an infinite amount of time $t$ can be simulated with a finite $\tau$.

As well as being useful in the single qubit case, these results can also be directly used in the case of a system of many qubits all subjected to independent channels, leading to a superpolynomial speed up from what could be achieved classically. Prominent questions which could be investigated include the decay of multipartite entanglement of an initially highly entangled state \cite{Levi2013}, or the performance of one-way computation \cite{Raussendorf2001} with a cluster state affected by local noise. For such systems of $N$ qubits the dilation can be calculated  once classically and the dynamics of the entire system then simulated on a quantum computer using a maximum of 2N ancilla qubits and Hamiltonians that affect, at most, 3 qubits. Simulating this classically would require applying the quantum channel up to $2^N$ times for an initial state which is highly entangled or, equivalently, solving the dynamics of the complete channel which would have $2^N$ Kraus operators.  

This method also provides the ability to do something which a normal Stinespring dilation cannot do at all. By simulating the dynamics continuously in time, the evolution of the state follows the `true' path that we are trying to simulate, rather than just reaching the required goal for a single instance in time. This means that the full information about the behaviour of the system over an extended interval of time is accessible, allowing simulations where the time at which measurement takes place is not known a priori.  In our scheme, such a scenario creates no additional difficulty, as the system follows the correct continuous dynamics. In a standard Stinespring dilation approach, however, this is either completely impossible or introduces substantial errors. These can be quantified by considering the snapshot Stinespring dilation as being a constant Hamiltonian (the logarithm of the unitary dilation) applied for a shorter or longer period of time, leading to an error which grows in time.

In our approach, the complexity of open system dynamics is condensed in the time-dependent system environment interaction, allowing a simulation to be implemented using state of the art methods; whereas in natural systems, and in previous approaches to simulating the dynamics of open systems, the complexity resides in the dynamics of the infinitely large environment. These two perspectives can be understood as the two ends of the spectrum of quantum simulations. Since any time-dependence can be understood as originating from the dynamics of an additional system (fundamentally all interactions are time-independent), our approach suggests very clearly how to access the entire spectrum: expand the ancilla system while gradually reducing the time-dependence of the interaction. Such a continuous variation will give valuable insight, for example in the controllability of open quantum dynamics, as seen in Eqs.(\ref{eq:drivenDila}) and (\ref{eq:RWAHamiltonian}). There the back action of an external control caused by the environment interaction makes itself transparent, whereas such effects are extremely hard to unravel in a model based on an infinitely large environment. The suggested transition would allow the study of this back-action in its entire range of manifestations, opening up a completely new angle on the investigation of control on open quantum systems. This, in turn, fosters our endeavours in the struggle against decoherence and the realisation of working quantum technologies.
\paragraph*{Acknowledgement}
We would like to thank Martin Fraas for fruitful discussions. This work was supported by EPSRC through the Quantum Controlled Dynamics Centre for Doctoral Training and the ERC project ODYCQUENT.

\bibliographystyle{apsrev4-1}
\bibliography{/Users/bd313/library.bib}

\appendix

\section{Appendix A: Completing the Unitary}
\label{sec:AppendixA}
In order to find a Hamiltonian for a dilation, we first have to find the corresponding unitary. As noted in Eq.(2), this unitary satisfies
\begin{align}
\bra{k_B}U(t)\ket{0_B} &= M_k(t) \;\;\forall\;k, \\
U(t)U^\dagger(t) &= \mathds{I}. \nonumber
\end{align}
We show that it is always possible to find a solution for $U(t)$ which is smooth whenever the $M_k(t)$ are smooth. The first condition constrains $d$ columns of the unitary as being orthonormal vectors and the second condition requires us to pick the remaining $d(R-1)$ columns such that they form a complete orthonormal set. At $t=0$ we use Gram-Schmidt from an arbitrary basis such that the none of the $d(R-1)$ columns are $0$. In order to ensure at later times that these extra columns vary continuously in time we use the vectors at $t$ as a basis for Gram-Schmidt at $t+\delta t$. Let the columns of $U(t)$ be expressed as the $Rd$ vectors $\ket{v_n(t)}$. At time $t+\delta t$ the first $d$ of these vectors are transformed according to the change of the Kraus operators. They are still orthonormal to each other, but no longer orthogonal to the other $R(d-1)$ vectors. Using Gram-Schmidt, we update the $d+1$ vector by first calculating:
\begin{align}
\label{gram1}
\ket{v'_{d+1}(t+\delta t)} &= \\
 \ket{v_{d+1}( t)} &- \sum_{n=1}^{n=d}\braket{v_{d+1}( t)|v_n( t+\delta t)}\; \ket{v_n( t+\delta t)} \nonumber.
\end{align}
For $n \le d$ we can also write
\begin{equation}
\ket{v_n( t+\delta t)} = \ket{v_n( t)} + \delta t\ket{\Delta_n( t)} + O(\delta t^2)
\end{equation}
whenever the $\ket{v_n( t)}$ are a smooth function of $t$, that is, whenever the $M_k(t)$ are smooth. This enables us to rewrite Eq.(\ref{gram1}) as
\begin{align}
\ket{v'_{d+1}(t+\delta t)} &= \nonumber\\
\ket{v_{d+1}( t)} - &\sum_{n=1}^{n=d}\delta t \braket{v_{d+1}( t)|\Delta_n( t)}\; \ket{v_n( t+\delta t)}+O(\delta t^2) \nonumber
\end{align}
due to the orthogonality at $ t$. This explicitly shows that $\ket{v'_{d+1}( t)}$ is a smooth function of $t$. From this we can easily obtain the normalised vector
\begin{align}
&\ket{v_{d+1}( t+\delta t)} =  \ket{v_{d+1}( t)}\\
 &+ \delta t \sum_{n=1}^{n=d} \text{Re}\left(\braket{v_{d+1}( t)|\Delta_n( t)} \braket{v_{d+1}( t)|v_n( t+\delta t)} \right)\ket{v_{d+1}( t)} \nonumber\\
 &- \delta t \sum_{n=1}^{n=d} \braket{v_{d+1}( t)|\Delta_n( t)}\; \ket{v_n( t+\delta t)} + O(\delta t^2)\nonumber
\end{align}
which is also a smooth function of $ t$ whenever the $M_k( t)$ are smooth. By induction, we see that this is true for all $\ket{v( t)}$ and all $t$. Hence we can always construct a unitary which is smooth whenever the Kraus operators are smooth.

\section{Appendix B: Divergence at $t=0$}
\label{sec:AppendixB}
We prove that the dilation of $\epsilon(t)$ diverges at $t=0$ if and only if $\dot{\epsilon}(0)$ is non-Hermitian. We first consider the case where the dilation does not diverge, which lets us write
\begin{align}
\epsilon_t(\rho) &= \text{Tr}_B[U(t)\rho\otimes\omega U^\dagger(t)] \nonumber \\
\dot{\epsilon}_t(\rho) &= \text{Tr}_B\left\{-i[H(t), U(t)\rho\otimes\omega U^\dagger(t)]\right\}\\
\dot{\epsilon}_0(\rho) &= -i \text{Tr}_B\left\{[H(0),\rho\otimes\omega]\right\} \nonumber
\end{align}
as $\epsilon_0(\rho) = \rho$ means that there exists a dilation such that $U(0)=\mathds{I}$. We expand $H(0)$ in terms of separable Hermitian operators:
\begin{equation}
H(0) = \sum_k A_k\otimes B_k.
\end{equation}
This enables us to calculate the partial trace according to
\begin{align}
\dot{\epsilon}_0(\rho)  &= -i \text{Tr}_B\left[ \sum_k\left(A_k\rho \otimes B_k\omega - \rho A_k\otimes \omega B_k\right) \right] \nonumber\\
&= -i \sum_k [\lambda_k A_k, \rho]\\
&= -i [H', \rho] \nonumber
\end{align}
where $\lambda_k = \text{Tr}(\omega B_k)$. These are necessarily positive as $B_k$ is Hermitian and $\omega$ is a state, from which it follows that $H'$ must also be Hermitian. This proves that if the dilation is finite at $t=0$, then $\dot{\epsilon}_0(\rho) $ must be purely Hermitian.

To prove the converse we recall that the divergence depends on the eigenvalues of the Choi state. Specifically, that at least one of the eigenvalues must obey $\lambda(0) = 0$ but $\dot{\lambda}(0) \ne 0$. The Choi state is pure if and only if the quantum channel is unitary. If this is the case it is clear that the eigenvalues do not change, so we ignore the Hermitian part of the channel. Assuming that $\dot{\epsilon}_0 = 0$, we can write the quantum channel for short times as
\begin{equation}
\epsilon_t \approx  \mathds{I} + X \frac{t^2}{2} + O(t^3).
\end{equation}
As previously stated, reshuffling the elements of $\epsilon_t$, written as a matrix, gives us the Choi state $\Lambda_t$. This means that the elements of $\Lambda_t$ have, potentially, terms of every order in $t$ \emph{apart} from first order. As the $\Lambda_t$ is a state, though not quite normalised, its eigenvalues are all non-negative and sum to $d$. It is also a pure state at $t=0$ and, as the eigenvalues are analytic in $t$, we can we write:
\begin{align}
\lambda_0 &= d + \alpha_0 t +\beta_0 t^2 + O(t^3) \nonumber\\
\lambda_i &= 0 + \alpha_i t +\beta_i t^2 + O(t^3)\;\;\; i=1,...,d^2-1
\end{align}
The eigenvalues are also obtained from the characteristic equation for $\Lambda_t$, and so we can write:
\begin{equation}
\lambda^{d^2} + A\lambda^{d^2-1} + B\lambda^{d^2-2}+... = 0
\end{equation}
where A, B, C,... are products and sums of the elements of $\Lambda_t$. This means that none of these coefficients can have terms which are linear in $t$. From the basic properties of the roots of polynomial equations we have that 
\begin{align}
B &= \sum_{\mu\neq\nu} \lambda_\mu \lambda_\nu\;\;\; \mu,\nu = 0,...,d^2-1\nonumber\\
&= d\sum_i\alpha_i t + O(t^2).
\end{align}
But, as we have already noted, B cannot have a term proportional to $t$. This implies that $\sum_i \alpha_i= 0$ and, as each are non-negative, that $\alpha_i$ = 0. The sum of the eigenvalues hence also means that $\alpha_0 = 0$. Therefore, all the eigenvalues obey $\dot{\lambda}(t=0)=0$ and, as noted previously, this is a sufficient condition for the dilation to not diverge.

\end{document}